\documentclass{sigchi}

\usepackage{booktabs} 
\usepackage{colortbl}
\usepackage{color}
\definecolor{gray}{rgb}{0.8,0.8,0.8}
\usepackage{gensymb}
\usepackage{multirow}
\usepackage{balance}       

\usepackage{balance}  
\usepackage{graphics} 
\usepackage{times}    
\usepackage{url}      
\usepackage{longtable}
\usepackage{caption}
\usepackage{subcaption}

\def\pprw{8.5in}
\def\pprh{11in}

\usepackage[pdftex]{hyperref}
\hypersetup{
pdftitle={SIGCHI Conference Proceedings Format},
pdfauthor={LaTeX},
pdfkeywords={SIGCHI, proceedings, archival format},
bookmarksnumbered,
pdfstartview={FitH},
colorlinks,
citecolor=black,
filecolor=black,
linkcolor=black,
urlcolor=black,
breaklinks=true,
}

\begin{document}

\title{Can Gaze Beat Touch? A Fitts' Law Evaluation of Gaze, Touch, and Mouse Inputs}


\numberofauthors{2}
\author{
\alignauthor Vijay Rajanna\\
       \affaddr{Sketch Recognition Lab. Dept. of Computer Science and Engineering}\\
       \affaddr{Texas A\&M University, College Station, Texas}\\
       \email{vijay.drajanna@gmail.com}
\alignauthor Tracy Hammond\\
       \affaddr{Sketch Recognition Lab. Dept. of Computer Science and Engineering}\\
       \affaddr{Texas A\&M University, College Station, Texas}\\
       \email{thammond@gmail.com}
}

\maketitle
\begin{abstract}
Gaze input has been a promising substitute for mouse input for point and select interactions. 
Individuals with severe motor and speech disabilities primarily rely on gaze input for communication.
Gaze input also serves as a hands-free input modality in the scenarios of situationally-induced impairments and disabilities (SIIDs).
Hence, the performance of gaze input has often been compared to mouse input through standardized performance evaluation procedure like the Fitts' Law.
With the proliferation of touch-enabled devices such as smartphones, tablet PCs, or any computing device with a touch surface, it is also important to compare the performance of gaze input to touch input.

In this study, we conducted ISO 9241-9 Fitts' Law evaluation to compare the performance of multimodal gaze and foot-based input to touch input in a standard desktop environment, while using mouse input as the baseline.  
From a study involving 12 participants, we found that the gaze input has the lowest throughput (2.55 bits/s), and the highest movement time (1.04 s) of the three inputs.
In addition, though touch input involves maximum physical movements, it achieved the highest throughput (6.67 bits/s), the least movement time (0.5 s), and was the most preferred input.
While there are similarities in how quickly pointing can be moved from source to target location when using both gaze and touch inputs, target selection consumes maximum time with gaze input.
Hence, with a throughput that is over 160\% higher than gaze, touch proves to be a superior input modality.
\end{abstract}

\section{Introduction}
\label{sec:fittsintro}
Advancements in ubiquitous computing is making computing capabilities available anywhere, anytime, and on any device~\cite{ubiquitous:charting_past_present_future}.
This also poses new challenges related developing new interaction methods beyond mouse, keyboard, and touch-based interactions, since a user's hands may not always be available to interact with a device.
For example, modern surgical technology has enabled a surgeon to view the imagery of the part of the body being operated in greater details through high resolution cameras.
However, if the surgeon wants to zoom, pan, switch images or perform any other interaction, she is required to go through the cumbersome process of changing the gloves, working on the computer, sterilizing the hands again, and putting on the glove.
These tasks would normally take around ten minutes that the surgeon cannot afford to lose~\cite{GazeTap:Benjamin}.
This is commonly referred to as situationally-induced impairments and disabilities (SIID)~\cite{Kane:walking}.
Other scenarios of situationally-induced impairments include, for example, a worker in a factory wearing gloves or having greasy hands that wants to perform some basic operations on a computer.
A person driving a car that wants to interact with the navigation map on a touch-enabled dashboard, etc.

While we discussed the needs for hands-free interactions in the context of situational impairments.
On the other hand, there are users with permanent disability or impairment that need hands free, accessible interactions.
Physically illiterate is a notion that indicates that a person with a disability faces challenges in acquiring mature movement patterns similar to their able-bodied peers~\cite{capio2011fundamental, whitehead2013definition}.
According to the 2016 disability status report (published in 2018) 12.8\% of the people in the United States have at least one kind of  disability~\cite{disabilitystatusreport2018}.
Hence, a hands-free method such as gaze-assisted interaction not only supports interactions in the scenarios of situational impairments, but also enables users with physical impairments to use the same interfaces and have the same experience as the able-bodied users when working on a computer.

Gaze-assisted interaction has already shown to support reliable point and select interactions similar to a mouse~\cite{Rajanna:GAWSCHI, Hansen:3D, Stellmach:looktouch, kumar:eyewindow} and support text entry through gaze typing~\cite{presstapflick:gazetyping, GazeTyping:adjustdwell:Majaranta, GazeTyping:GazeTalk:Hansen}.
Also, gaze input can be used as an accessible and a secure authentication method against shoulder-surfing attacks~\cite{Khamis:switch, IEEE:biometrics, Khamis:gazetouchpin}, and gaze input supports gestural interactions~\cite{amy:gazegesture, Vidal:pursuits}.
Gaze input is increasingly used as a substitute to conventional mouse and keyboard-based input method to support accessible interactions~\cite{duchowski2007eye, Gazetyping:twentyYears:Majaranta}, and gaze input enables advanced interactions in the scenarios of situationally-induced impairments and disabilities~\cite{GazeTap:Benjamin, IEEE:biometrics}.
Hence, it is essential to evaluate how does gaze input compare to some of the common input methods such as mouse and touch, and this motivated us to conduct Fitts' Law evaluation.

With gaze-assisted interaction, a user points at the target element with her gaze, and either dwelling for a specified time~\cite{jacob1991use, dwelltime:jacob1993} or blinking~\cite{blink:biswas, blink:motordisability} is used to select the target.
However, dwell or blink-based selection is known to result in usability issues, reduced performance, high error rate, and visual fatigue~\cite{GazeTyping:adjustdwell:Majaranta, isokoski2000text, Urbina2010}.
Recent works have demonstrated that gaze input can achieve better usability, lower error rate, and comparable performance to a mouse when using multimodal gaze input~\cite{Rajanna:GAWSCHI, manukumar:keyselect, gazeandfoot:Klamka2015, presstapflick:gazetyping}.
Importunately, with multimodal gaze input target selection is instantaneous instead of dwelling, and inadvertent selections are eliminated.
In this mode, gaze is primarily used for pointing and a secondary input is used for selection.
Multimodal gaze-based inputs such as gaze and speech~\cite{GazeTyping:Beelders2012}, gaze and foot~\cite{Rajanna:GAWSCHI, GazeTap:Benjamin}, and gaze and touch~\cite{chandan:tagswipe} are some of the examples.
Hence, in our experiment, we used a multimodal gaze and foot-based framework for gaze input~\cite{Rajanna:GAWSCHI}.

\subsection{An Overview of Fitts' Law Evaluation}
As non-keyboard input methods are developed, Fitts' law is often used in HCI research to quantify human performance~\cite{mackenzie:fittstool} on such novel input devices.
Fitts' Law models the human movement analogous to the way information is transmitted~\cite{SOUKOREFF2004751}.
Different kinds of movement tasks have different indices of difficulties expressed in bits/s.
To perform a movement task, a certain number of bits of information is transmitted by the human motor system.
The performance of a movement task can be quantified (throughput) by dividing the number of bits transmitted by the movement time (MT)~\cite{SOUKOREFF2004751}.
Furthermore, Fitts' Law has been used in HCI research in two ways, first, to predict the time it takes (movement time) for a user of a graphical interface to move the cursor to the target and click it.
Second, to compare the speed and accuracy of different input methods through a single statistic called throughput~\cite{mackenzie:fittstool, SOUKOREFF2004751, Zhang2007}.
The throughput of an input method is computed as shown in Equation~\ref{eqn:tp}.

\begin{equation}
\label{eqn:tp}
Throughput = \frac{ID_{e}}{MT}
\end{equation}

Where, $ID_{e}$ is the effective Index of Difficulty, and MT is the mean Movement Time.
The subscript ${e}$ indicates ``effective." While $ID$ represents the Index of Difficulty considered for the tasks, in its effective form, i.e., $ID_{e}$, represents the difficulty of the task completed by the user rather than the what she was presented with~\cite{SOUKOREFF2004751}.
The $ID_{e}$ is calculated as shown in Equation~\ref{eqn:ide}.

\begin{equation}
\label{eqn:ide}
    ID_{e} = log_{2}\big(\frac{A_{e}}{W_{e}}+1\big)
\end{equation}

Where, $A_{e}$ is the effective distance to the target (amplitude), i.e., the mean of the effective movement amplitudes measured along the task axis over a sequence of trials.
The effective amplitude of a trial in a sequence is computed as $A + dx$, and the $dx$ is computed as shown in Equation~\ref{eqn:dx}.
$W_{e}$ is the effective target width which is calculated as shown in Equation~\ref{eqn:we}.

\begin{equation}
\label{eqn:we}
  W_{e} = 4.133 \times SD_x 
\end{equation}

Where, $SD_x$ is the standard deviation of the selection coordinates ($dx$  - overshoot or undershoot) in a sequence of trials which is computed as shown in Equation~\ref{eqn:dx}~\cite{mackenzie:throughput_touch}.

\begin{equation}
\label{eqn:dx}
dx = (C*C - B*B - A*A)/(2.0*A)
\end{equation}

\begin{figure}[!ht]
  \includegraphics[width=3.4in]{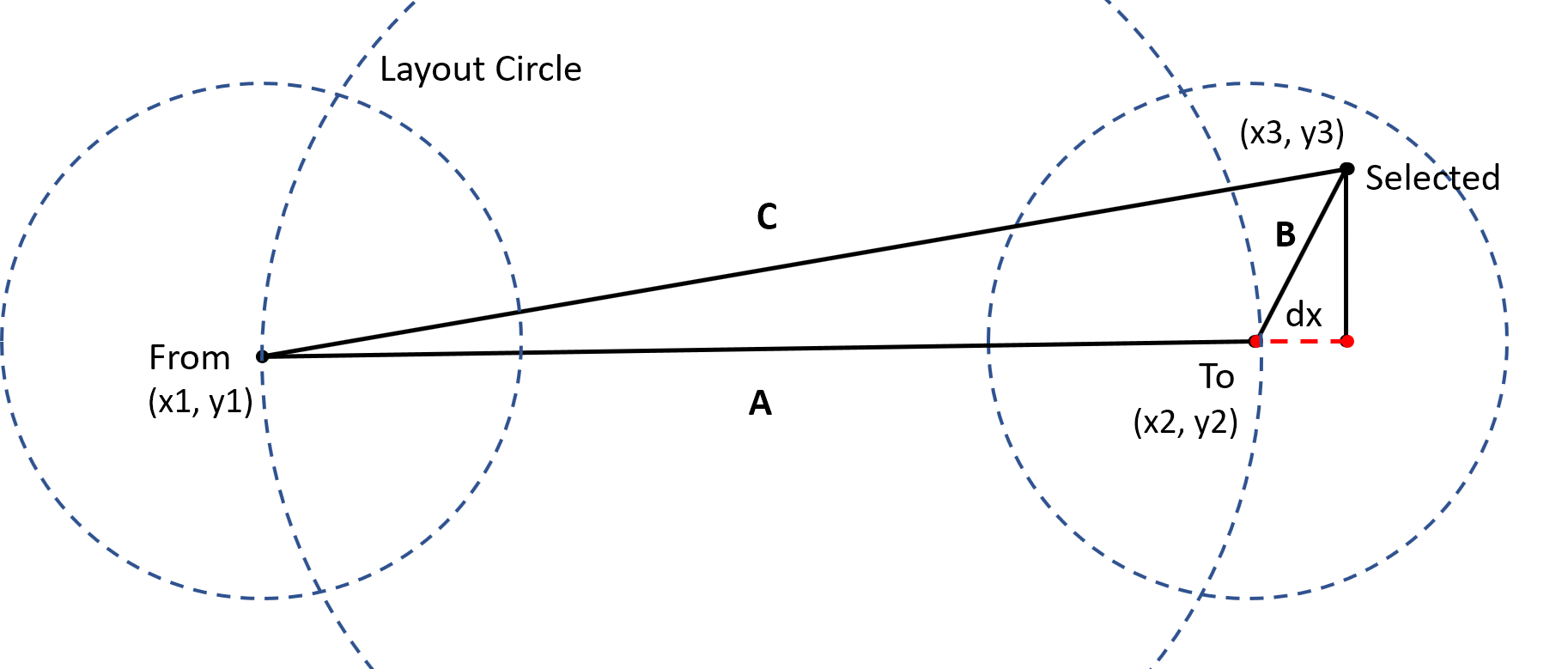}
  \caption{Computation of $dx$ used in the calculation of the Effective Amplitude $A_{e}$ and Effective Target Width $W_{e}$. The amount by which the user overshoots or undershoots from the center of the target is projected back on to the task axis. This a) ensures the inherent one-dimensionality of Fitts' Law, and b) the difficulty of the task is computed based on the actual task completed by the user rather than what she was presented to do.}
  \label{fig:fittserrorcalc}
\end{figure}

To compute $dx$, first the selection coordinate of a trial is projected back on to the task axis as shown in Figure~\ref{fig:fittserrorcalc}, and this is done to maintain the inherent one-dimensionality of Fitts' Law~\cite{SOUKOREFF2004751, Zhang2007}.
The task axis is the line joining the center of source (from) to destination target (to).
The values of A, B, and C are computed as shown in Equations~\ref{eqn:A},~\ref{eqn:B},~\ref{eqn:C}~\cite{mackenzie:throughput_touch}.

\begin{equation}
\label{eqn:A}
A = \sqrt{(x_2-x_1)^2 + (y_2-y_1)^2}
\end{equation}
\begin{equation}
\label{eqn:B}
B = \sqrt{(x_3-x_2)^2 + (y_3-y_2)^2}
\end{equation}
\begin{equation}
\label{eqn:C}
C = \sqrt{(x_1-x_3)^2 + (y_1-y_3)^2}
\end{equation}

\section{Prior Work}
\label{sec:fittspriorwork}
 Fitts' Law evaluation of gaze input has been conducted by primarily using gaze and dwell-based selection~\cite{Zhang2007, Ware:Input2, Vertegaal:Roel}.
However, multiple research works have conducted Fitts' Law evaluations by combining gaze with a secondary input in a multimodal gaze input setup~\cite{Vertegaal:Roel, Beelders:gazespeech, surakka2004gazing}.

Zhang et al., presented the first work on Fitts' law evaluation of gaze input that conforms to ISO 9241-9~\cite{Zhang2007}.
The authors compared gaze input with short and long dwell times and gaze+Spacebar with mouse input.
The target widths chosen were 75 px and 100 px, and amplitude chosen were 275 px.
The Gaze+Spacebar eliminated the waiting time, it was the best selection method among gaze inputs with a throughput of 3.78 bits/s (mouse was 4.68 bits/s).
Also, the participants liked Gaze+Spacebar out of the three gaze-based inputs.

Ware et al., presented a Fitts' law evaluation of gaze input~\cite{Ware:Input2}.
Three selection methods were used along with gaze: a button press, dwelling, and an onscreen select button.
In an experiment where the participants had to select one of the seven menu items arranged vertically, each item covering a visual angle of 2.0 to 1.65 degrees, the authors found that irrespective of the selection procedure, the gaze-based selection methods took less than 1 second for target selection.
Also, target selection with eye movements fits the Fitts' law well.

Miniotas et al., tested the validity of the findings from Ware et al.~\cite{Ware:Input2}, by comparing the performance of an eye tracker and a mouse in a simple pointing task~\cite{Miniotas:eyeandmouse}.
The participants had to make rapid and accurate horizontal movements to targets that were vertical ribbons.
The target amplitude included 26, 52, 104, and 208 mm and the target widths included 13 and 26 mm.
The authors found that the selection time is longer for the eye tracker than for the mouse by a factor of 2.7.

Zhai et al.~\cite{Zhai:1999:MGI}, proposed MAGIC: Manual and Gaze Input Cascaded Pointing to improve the usability of gaze input.
A Fitts' law evaluation was conducted by using 3 input methods which included an isometric pointing stick and two versions of MAGIC pointing.
The experiment included two target sizes (20, 60 px) and three target distances (200, 500, and 800 px).
The authors found that the completion time and target distance did not completely follow Fitts' law when using MAGIC pointing, but when considering both target size and target distance the data fit the Fitts' law but relatively poorly.
Pointing with two version of MAGIC achieved a higher performance (4.55 and 4.76 bits/s) than manual input (3.2 bits/s).


Vertegaal et al.~\cite{Vertegaal:Roel}, evaluated 4 input methods in a Fitts' task involving large visual tasks.
The input methods included a gaze and manual click, a gaze and dwell click, a stylus, and a mouse.
Unlike the Fitts' law task in~\cite{MacKenzie:ISO_9241Part9} that used ISO multi-directional tapping task, the authors in this experiment aimed at using gaze input to disambiguate between contexts of interaction, e.g., selecting one of the two large windows on a screen.
Hence, the experiment involved alternate selection of one of the two large visual targets (tasks with low index of difficulty).
The target widths included 70 px, 100 px, and 140px, and the amplitudes included 200 px, 400 px, and 800 px.
The index of difficulty varied from 1.28 bits/s to 3.6 bits/s.
In this experiment, gaze-based inputs outperformed manual input methods: mouse and stylus achieved an index of performance (IP corrected) of 4.7 and 4.2 bits/s respectively, but gaze with manual click and gaze with dwell (100 ms) achieved an IP of 10.9 and 13.8 bits/s respectively.
Though gaze input outperformed manual input, it also had higher error rate: mouse 4.6\%, stylus 6.2\%, gaze with manual click 11.7\%, and gaze with dwell click 42.9\%.
The authors concluded that gaze input with manual click provides the best trade-off between speed and accuracy.

Bleeders et al.~\cite{Beelders:gazespeech}, evaluated three gaze+speech inputs against a mouse in an ISO multi-directional tapping task.
The three gaze+speech based inputs included eye gaze and speech (ETS), ETS with magnification (ETSM), and ETS with gravitational well (ETSG).
The authors found that the mouse was far superior in performance when selecting the targets (throughput), compared to all gaze-based inputs.

Surakka et al.~\cite{surakka2004gazing} compared target acquisition of gaze pointing and EMG selection (i.e., frowning) to the mouse. The mouse was most effective for short distances, but the gaze+EMG input combination showed a higher index of performance than the mouse for error-free data, suggesting that gaze+EMG may be faster at longer distances, but their data did not show any speed advantage of gaze+EMG over the mouse. San Agustin et al.~\cite{san2009evaluation} later confirmed, that gaze+button and gaze+EMG were in fact faster than mouse+button and mouse+EMG.

The first ISO 9241-9 Fitts' Law evaluation of gaze input in virtual reality was conducted by Hensen et al.~\cite{cogain:fitts_hansen}.
Gaze pointing was compared with head pointing, while considering mouse input as baseline for both conditions.
A dwell time of 300 ms and clicking were used as two selection methods for all three inputs.
Gaze pointing and head pointing achieved comparable throughputs of 2.1 bits/s and 2.5 bits/s respectively.
In a follow up Fitts' Law evaluation, Hensen et al.~\cite{etra:fitsvr_foot} compared gaze, foot, and head pointing in virtual reality.
Gaze input was found to be the slower compared to other pointing methods.
Also, throughputs for gaze (2.55 bits/s) and foot (5.58 bits/s) pointing were lower than mouse (3.87 bits/s) and head (3.4 bits/s) pointing and their effective target widths were also higher.

Considering Fitts' Law evaluation of touch input, MacKenzie~\cite{mackenzie:throughput_touch} demonstrated the superiority of direct pointing device, i.e., touch input with a finger through Fitts' experiment.
The experiment involved 16 participants where they performed 1D and 2D Fitts tasks.
Overall, touch-based target selection achieved the throughput of 6.95 bits/s, which is significantly (50\%) higher than than accepted values for a mouse (3.7 bits/s to 4.9 bits/s).

A study which is similar to the current work was conducted by Rajanna et al.~\cite{fitslarge}, who compared gaze input to touch input on a large display (84-inch).
The work was motivated by the fact that in the scenarios of situationally-induced impairments, it is more convenient to use gaze-based multimodal input than touch or mouse inputs.
From a study involving 23 participants, it was found that gaze input had the lowest throughput (2.33 bits/s) and highest movement time (1.8 s).
Despite the large display size, touch input still achieved highest throughput of 5.49 bits/s and least movement time of 0.6 s.
The poor performance of gaze input was attributed to lower tracking accuracy of the eye tracker on the large display (error average  4.6\degree, max 9.0\degree).

While gaze input can move quickly from source to target location through saccadic eye movements~\cite{duchowski2007eye}, this raises a question if gaze can beat the performance of direct input method such as touch.
As the advancements in eye tracking hardware and algorithms resulted in lower gaze tracking error (0.4\degree to 0.9\degree deg), it is essential to compare the performance of gaze input to touch in a desktop setting (20-inch to 30-inch display).

\section{Fitts' Law Experiment Design and Evaluation}
\begin{figure}[!ht]
  \centering
  \includegraphics[width=2.5in]{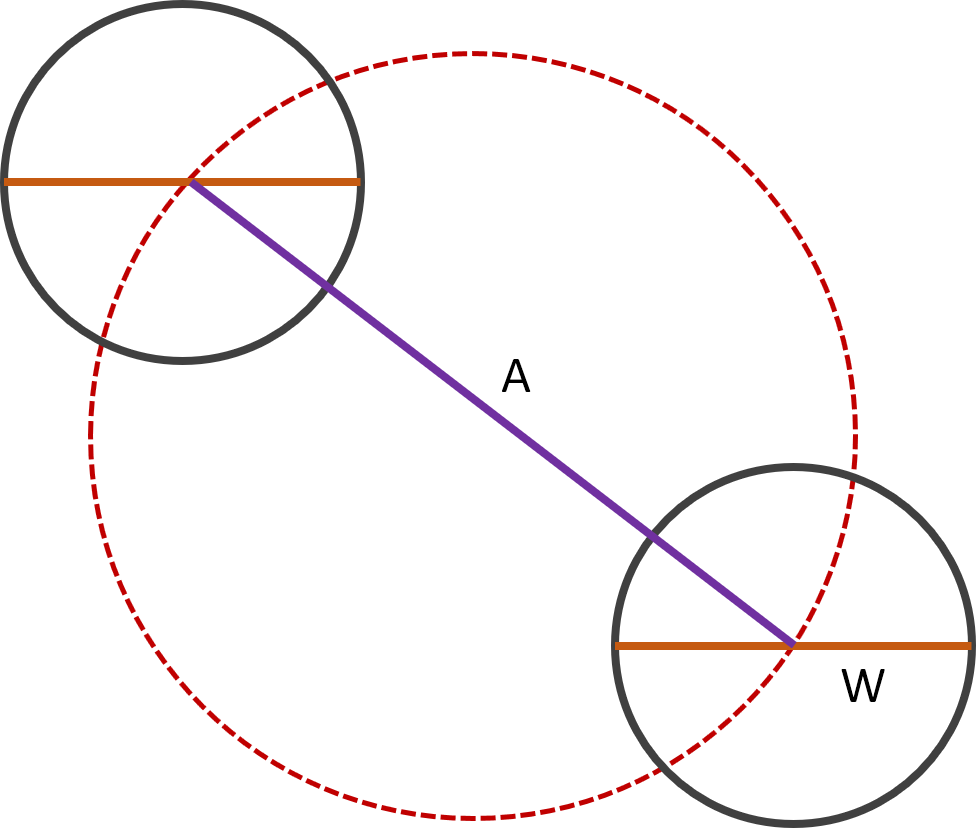}
  \caption{Fitt's Law Task Setup: N number of target circles with a width ``W" pixels in diameter are arranged around a layout circle with a width (amplitude) ``A" pixels in diameter.}
  \label{fig:fitts_target_arrangement}
\end{figure}

For the Fitts' Law experiment we used the software~\footnote{http://www.yorku.ca/mack/FittsLawSoftware/ [last accessed Jan 23rd 2022]} developed by Soukoreff and MacKenzie~\cite{SOUKOREFF2004751, mackenzie:fittstool}.
Specifically, we used Fitts' Task Two which is a multi-directional point-and-select task with targets arranged along the periphery of a layout circle with diameter A and the width (diameter) of each target was W as shown in Figure~\ref{fig:fitts_target_arrangement}.

\begin{figure}[!ht]
  \centering
  \includegraphics[width=3.3in]{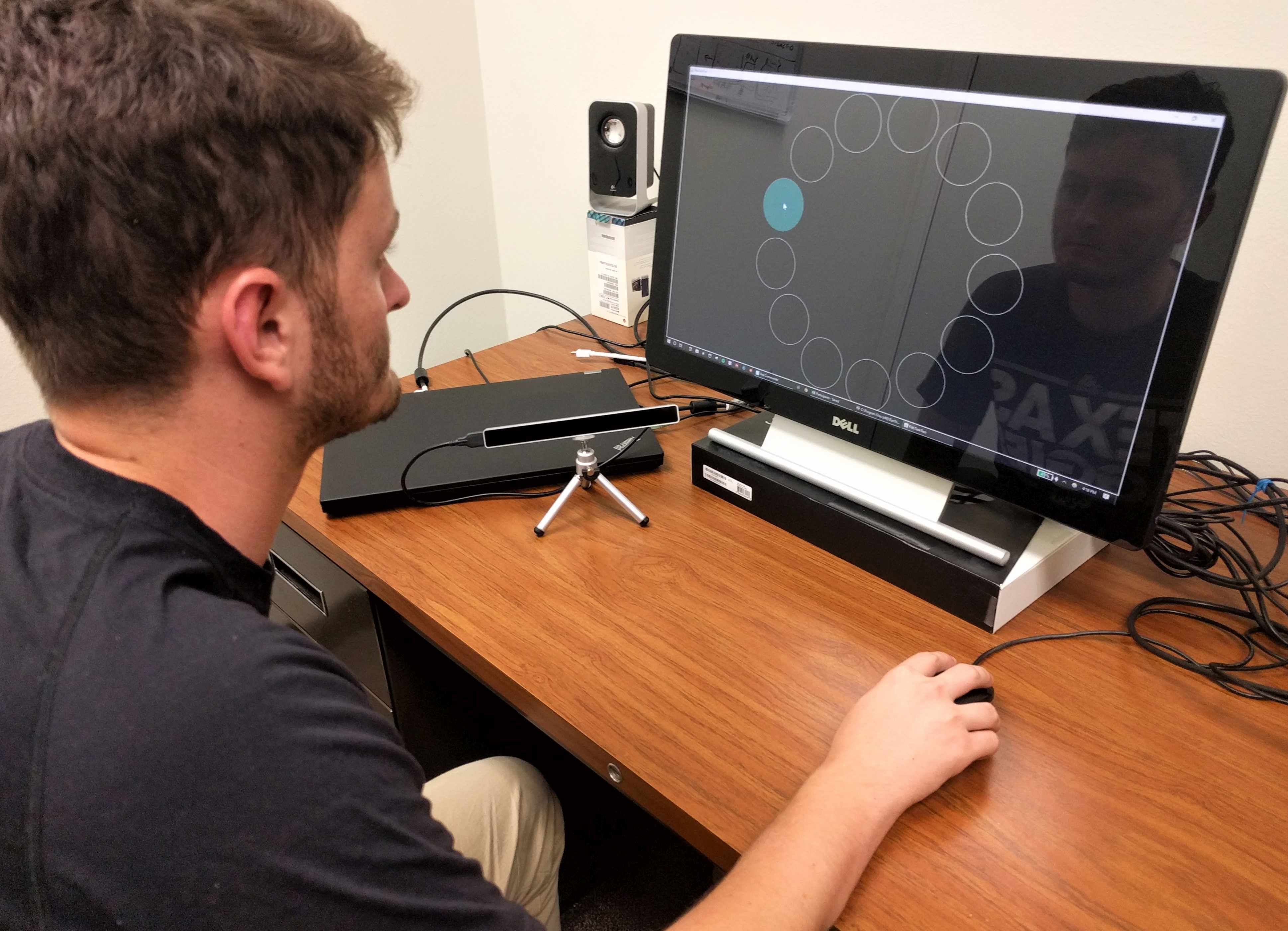}
  \caption{Fitts' Law Evaluation: mouse input}
  \label{fig:fitts_smallMouse}
\end{figure}

For each trial the target to be selected is highlighted in red color, and once the highlighted target is selected, the target that is opposite to the current target gets highlighted.
In accordance with the previous Fitts' law studies on gaze pointing, we used a nominal index of difficulty that ranged from 2.0 to 2.5~\cite{Zhang2007}.
Hence, the amplitude, i.e., the distance to the target we chose were 1000 px and  1100 px, and the target widths were set to 230 px and 330 px.
The computation of the index of difficulty is shown in Table~\ref{table:standardscreenID}.
Figure~\ref{fig:fitts_smallMouse} shows the experiment setup where the Fitts' Law task is shown on a standard 24-inch display.

\setlength\tabcolsep{3.5pt} 
\begin{table}[!ht]
\centering
\caption{Fitts' Law Evaluation - Standard 24-inch display: Amplitude, Width, and Index of Difficulty }
\label{table:standardscreenID}
\def\arraystretch{1.8} 
\begin{tabular}{ccc}
\hline
\rowcolor{gray} 
\textbf{Amplitude (px)} & \textbf{Width (px)} & \textbf{Index of Difficulty (bits/s)} \\ \hline
1100 & 230 & 2.53 \\ \hline
\rowcolor{gray}1000 & 230 & 2.41 \\ \hline
1100 & 330 & 2.11 \\ \hline
\rowcolor{gray}1000 & 330 & 2.01 \\ \hline
\end{tabular}
\end{table}

\subsection{Selection Methods}
\begin{figure}[!ht]
  \centering
  \includegraphics[width=3.4in]{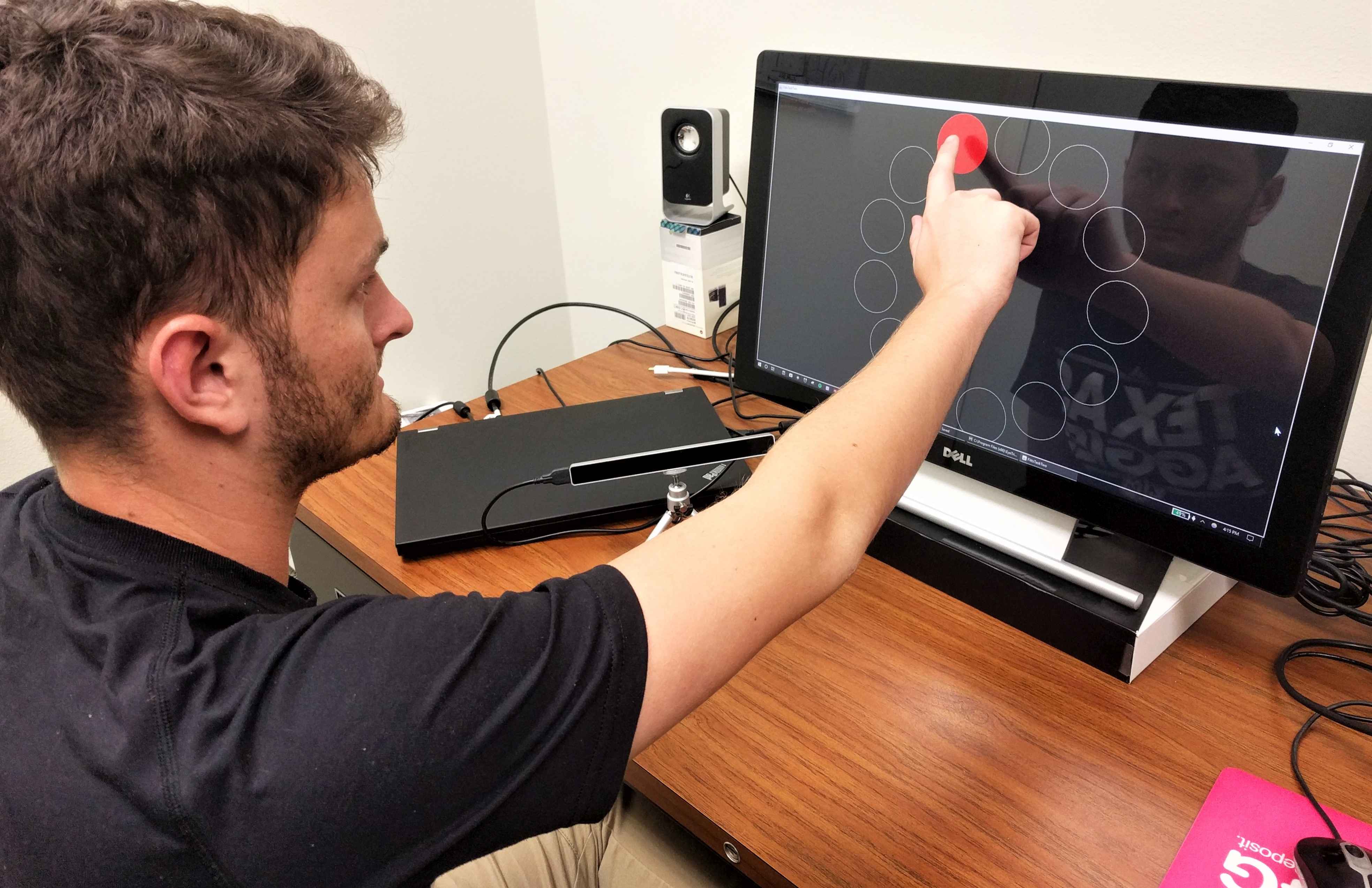}
  \caption{Fitts' Law Evaluation: touch input}
  \label{fig:fitts_smallTouch}
\end{figure}

\begin{figure}[!ht]
  \centering
  \includegraphics[width=3.4in]{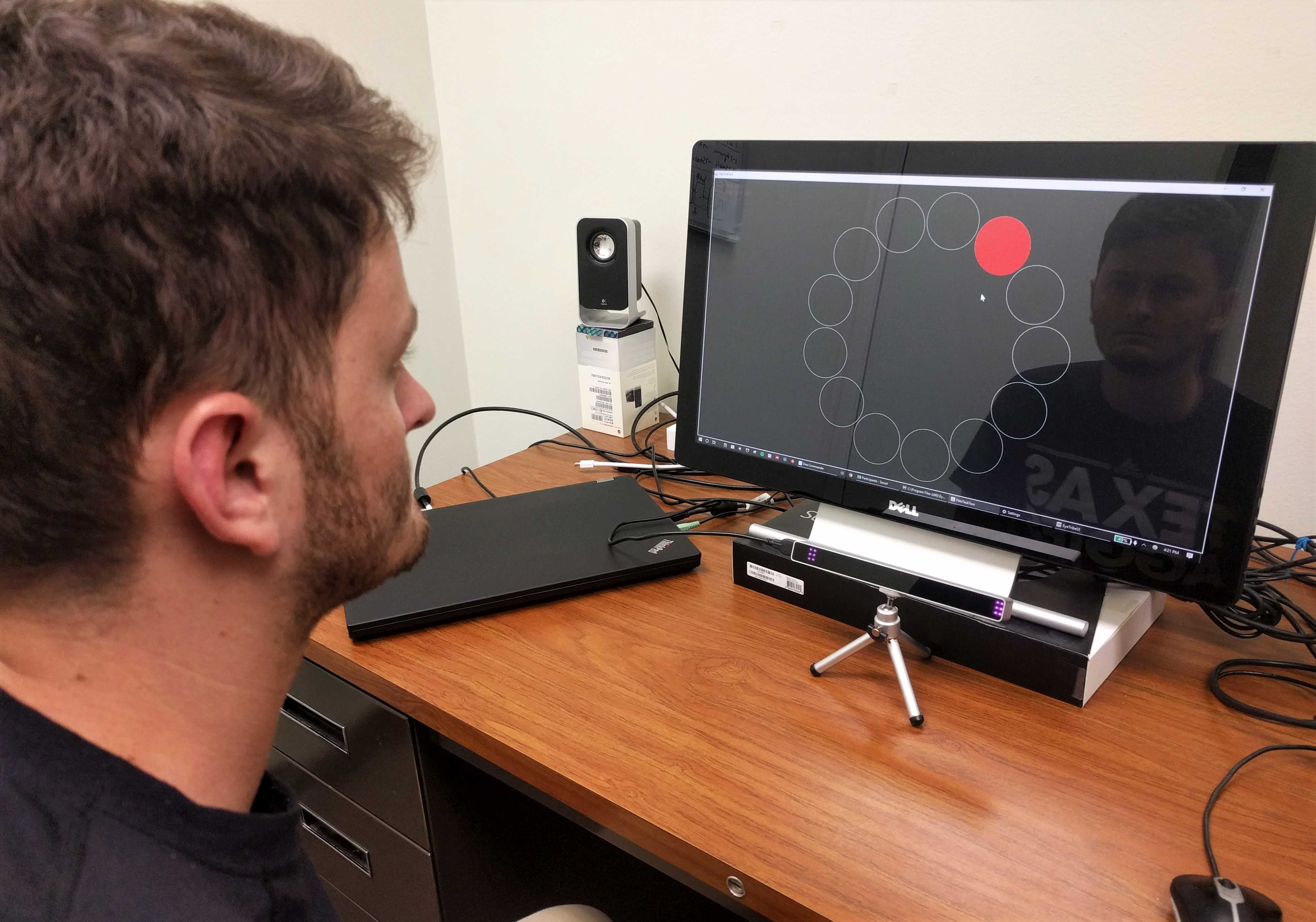}
  \caption{Fitts' Law Evaluation: gaze input}
  \label{fig:fitts_smallgaze}
\end{figure}

We chose three selection methods: 1) mouse, 2) touch, and 3) gaze+foot.
For the mouse input, the participant used a standard mouse to select targets as shown in Figure~\ref{fig:fitts_smallMouse}.
The cursor speed was set to the default value.
For the touch input, the participant directly touched the screen to select targets as shown in Figure~\ref{fig:fitts_smallTouch}.
For gaze+foot input an eye tracker was placed in between the user and the display as shown in Figure~\ref{fig:fitts_smallgaze}.
The eye tracker was removed when using the mouse and touch inputs.

\begin{figure}[!ht]
  \centering
    \includegraphics[width=3.38in]{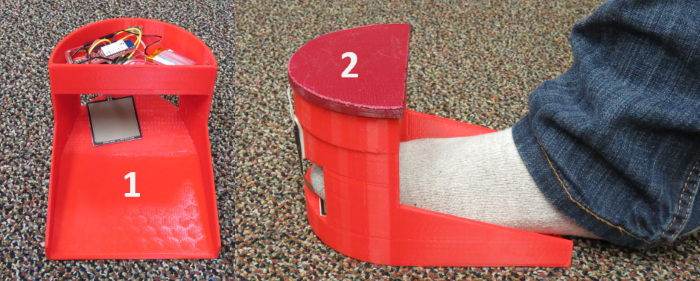}
  \caption{The foot controller used in the gaze+foot selection method. 1 - a force sensitive resistor, microcontroller, and bluetooth module in a 3D printed case, 2 - foot interaction.}
  \label{fig:foot}
\end{figure}

To achieve the gaze+foot interaction, we enhanced a gaze+foot input system developed by Rajanna et al.~\cite{Rajanna:GAWSCHI}, which consists of an eye tracking module and a foot controller (Figure~\ref{fig:foot}), and the on-screen cursor follows the user's gaze.
To select a target the user first places the cursor on the target by focusing on it, and then selects it by pressing a pressure sensor, attached to the foot controller, with the foot.
The foot controller connects to the eye tracking system over Bluetooth, and the entire circuitry is placed inside a portable 3D printed case.

\subsection{Display and Gaze Tracking}
The experiment was conducted on a Dell Monitor, a 24" touch enabled display.
We used an Eye Tribe tracker for eye tracking.
The tracker had a manufacturer reported accuracy of 0.5\degree~to 1.0\degree~of visual angle, and had a sampling rate of 60 Hz.

\subsection{Participants and Procedure}
For the Fitts' Law experiment we recruited 12 participants (8 M, 4 F) with their ages ranging from 19 to 32 ($\mu_{age}=23$).
At the beginning of the study, each participant was briefed about the Fitts' Law task and the kind of inputs they would be using for target selection.
For each input method (e.g., mouse) the participant completed one sequence of trials to familiarize themselves with the system before the actual data collection began.
The participants used three input methods--gaze+foot, mouse, and touch--for target selection, and the order of input methods used by the participants was counterbalanced according to the Latin square design.

For each input method the participant completed 4 blocks of target selection task, and each block had four sequences of trials as we used two amplitudes (1000 px and  1100 px) and two target widths (230 px and 330 px).
In each sequence, there were 13 trials, hence, a total of 2,496 trials (13 trials $\times$ 4 seq $\times$ 4 blocks $\times$ 12 participants) were completed for each input.
Also, a total of 7,488 trials (2,496 $\times$ 3 inputs) were completed from all the three inputs.
The participants were allowed to rest for a minute between each block, and in the case of gaze input, the participants were re-calibrated if the calibrated stance was disturbed between the blocks.

\section{Results}
\label{sec:fittssmallresult}
We conducted a one-way ANOVA with replication on the four dependent variables (DVs): 1) movement time, 2) throughput, 3) error rate, and 4) effective target width.
The independent factor was the `selection method' which had three levels: 1) mouse, 2) touch, and 3) gaze.
Table~\ref{table:fittsAnovasmallscreen} shows the result of ANOVA on the DVs, and also the mean and standard deviation of the selection methods for each DV.

\setlength\tabcolsep{1.5pt} 
\begin{table}[!ht]
\centering
\caption{Fitts' Law Evaluation: ANOVA and post-hoc analysis (p values highlighted in gray indicate significance at $\alpha=0.001$).}
\label{table:fittsAnovasmallscreen}
\def\arraystretch{1.8} 
\begin{tabular}{llll}
\cline{2-4}
\begin{tabular}[c]{@{}l@{}}Selection Method\\ {[}Ms, Th, Gz{]}\end{tabular} & \multicolumn{1}{c}{\cellcolor{gray}\textbf{Mean}} & \cellcolor{gray}\textbf{Std. Dev} & \multicolumn{1}{c}{\cellcolor{gray}\textbf{ANOVA}} \\ \hline
\cellcolor{gray}\textbf{\begin{tabular}[c]{@{}l@{}}Movement \\ Time (ms)\end{tabular}} & \begin{tabular}[c]{@{}l@{}}Ms = 683.9\\ Th = 490.7\\ Gz = 1040.4\end{tabular} & \begin{tabular}[c]{@{}l@{}}148.5\\ 101.6\\ 294.8\end{tabular} & \begin{tabular}[c]{@{}l@{}}F(2,382) = 633.1\\ \colorbox{gray}{\textit{p $<$ 0.001}}\end{tabular} \\ \hline

\cellcolor{gray}\textbf{\begin{tabular}[c]{@{}l@{}}Throughput\\ (bits/s)\end{tabular}} & \begin{tabular}[c]{@{}l@{}}Ms = 3.8\\ Th = 6.6\\ Gz = 2.5\end{tabular} & \begin{tabular}[c]{@{}l@{}}0.8\\ 1.5\\ 0.9\end{tabular} & \begin{tabular}[c]{@{}l@{}}F(2,382) = 797.8\\ \colorbox{gray}{\textit{p $<$ 0.001}}\end{tabular} \\ \hline

\cellcolor{gray}\textbf{Error Rate (\%)} & \begin{tabular}[c]{@{}l@{}}Ms = 1.3\\ Th = 0.3\\ Gz = 3.0\end{tabular} & \begin{tabular}[c]{@{}l@{}}3.6\\ 1.5\\ 5.1\end{tabular} & \begin{tabular}[c]{@{}l@{}}F(2,382) = 26.4\\ \colorbox{gray}{\textit{p $<$ 0.001}}\end{tabular} \\ \hline

\cellcolor{gray}\textbf{\begin{tabular}[c]{@{}l@{}}Effective Target\\ Width (pixels)\end{tabular}} & \begin{tabular}[c]{@{}l@{}}Ms = 229.7\\ Th = 148.1\\ Gz = 304.5\end{tabular} & \begin{tabular}[c]{@{}l@{}}124.5\\ 149.2\\ 327.8\end{tabular} & \begin{tabular}[c]{@{}l@{}}F(2,382) = 23.8\\ \colorbox{gray}{\textit{p $<$ 0.001}}\end{tabular} \\ \hline

\end{tabular}
\end{table}

\begin{figure}[!h]
  \centering
  \includegraphics[width=2.5in]{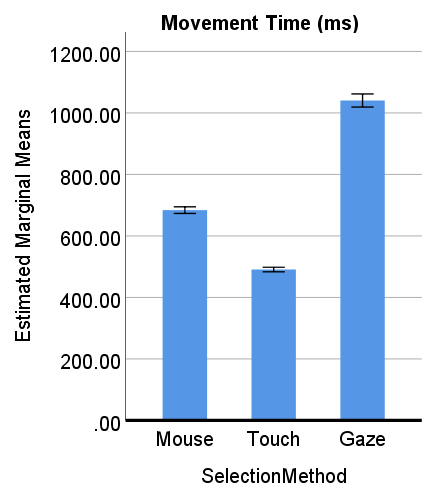}
  \caption{Movement time: comparison of movement time between the three selection methods (lower value is better).}
  \label{fig:mtsmall}
\end{figure}

\begin{figure}[!h]
  \centering
  \includegraphics[width=2.5in]{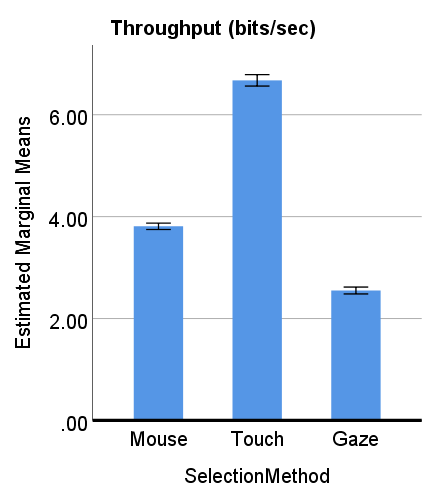}
  \caption{Throughput: comparison of throughput between the three selection methods (higher value is better).}
  \label{fig:tpsmall}
\end{figure}

\begin{figure}[!h]
  \centering
  \includegraphics[width=2.5in]{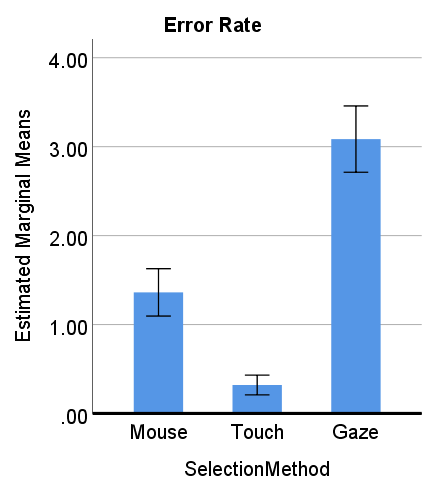}
  \caption{Error rate: comparison of error rate between the three selection methods (lower value is better).}
  \label{fig:ersmall}
\end{figure}

\begin{figure}[!h]
  \centering
  \includegraphics[width=2.5in]{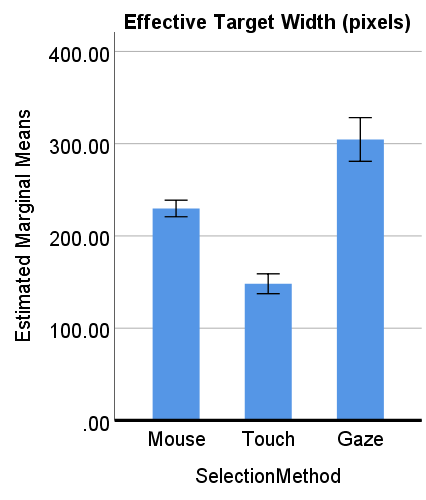}
  \caption{Effective target width: comparison of effective target width between the three selection methods (lower value is better).}
  \label{fig:wesmall}
\end{figure}

We observe that the factor `selection method' is significant (p $<$ 0.001) for all the four DVs, i.e., the value of a DV differs among the selection methods.
Out of all the selection methods, `touch' achieves the highest throughput (6.67 bits/s), consequently it has the least movement time, error, and effective target width.
Similarly, `gaze' input has the lowest throughput (2.55 bits/s), consequently it has the highest movement time, error, and effective target width.
Post-hoc tests with Bonferroni correction showed that for DVs movement time, throughput, and effective target width the difference between each pair of the selection methods, (mouse, touch) (mouse, gaze) (touch, gaze), was significant (p $<$ 0.05).
Figure~\ref{fig:mtsmall}, Figure~\ref{fig:tpsmall}, Figure~\ref{fig:ersmall}, and Figure~\ref{fig:wesmall} compare the means of the four DVs across the three selection methods.

\section{Discussion}
\label{sec:fitsssmalldiscussion}
From the results in Table~\ref{table:fittsAnovasmallscreen} we observe that touch input achieves the highest throughput, and it has the least movement time, error rate, and effective target width.
Similarly, gaze has the lowest throughput, highest movement time, error rate, and effective target width.
These results reflect that direct manipulation (input) method like touch is the fastest and most accurate input technique.

\begin{figure}[!ht]
  \centering
  \includegraphics[width=3.0in]{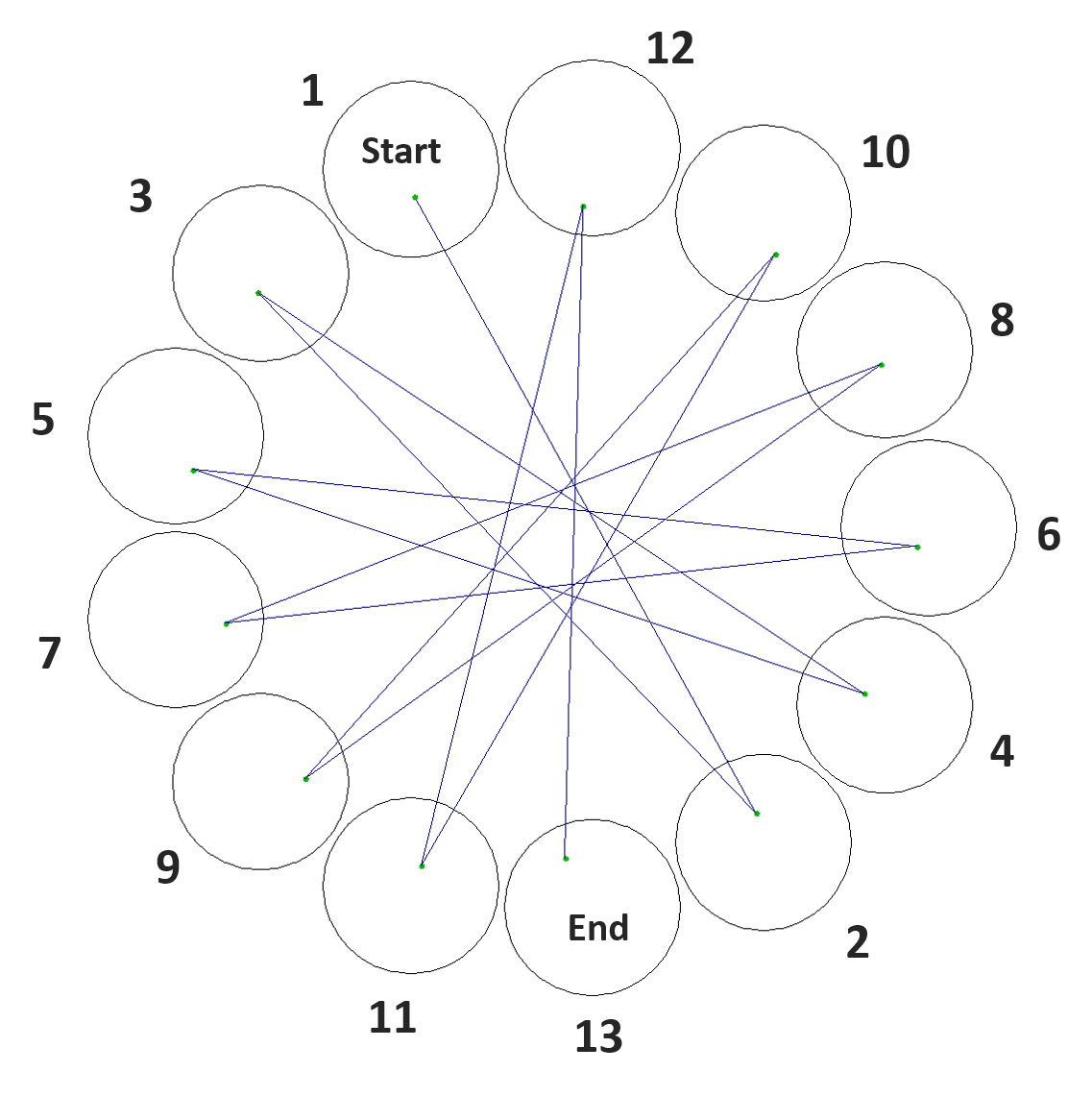}
  \caption{Touch input cursor points visualization: by observing transition paths, it is evident that each transition is well planned and hence the user hardly overshoots the target. Also, selections are closer to the center of each target resulting in lower effective target width and error.}
  \label{fig:touchvisualization}
\end{figure}

\begin{figure}[!ht]
  \centering
  \includegraphics[width=3.0in]{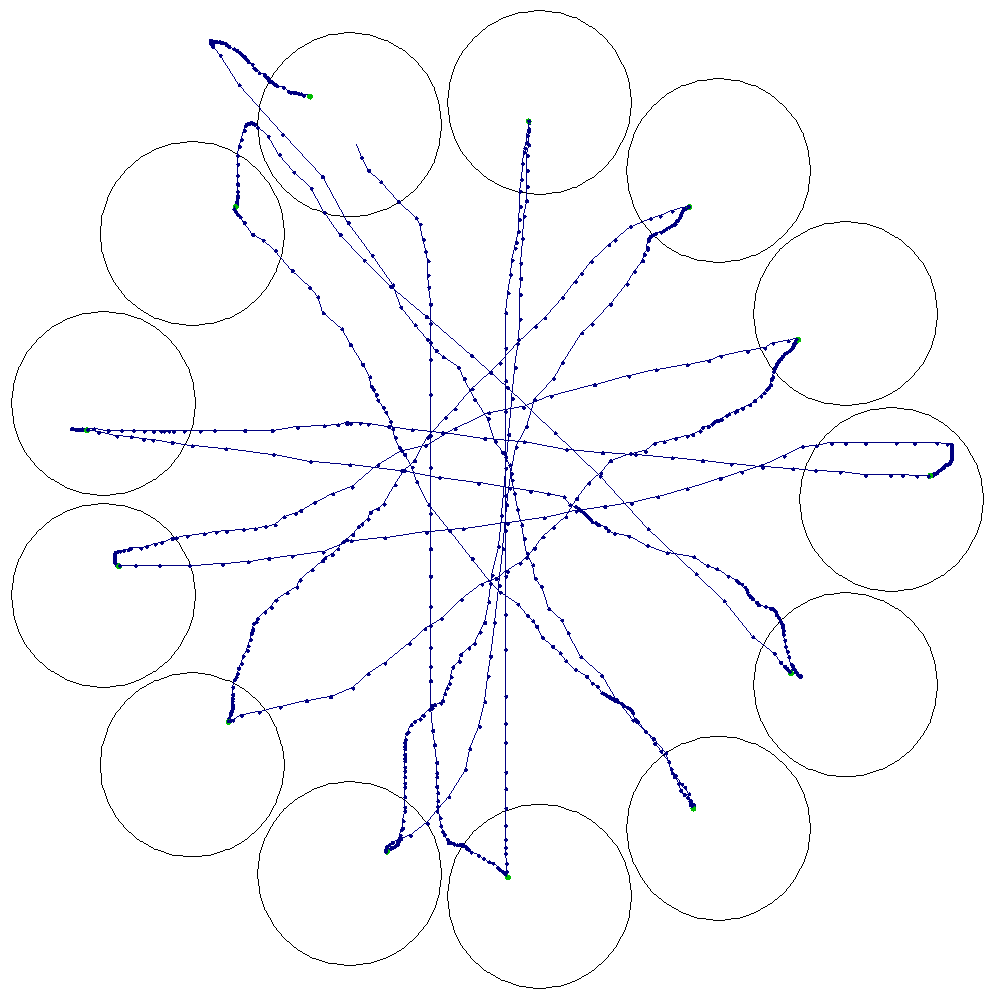}
  \caption{Mouse input cursor points visualization: out of the three inputs, mouse input has the maximum number of cursor points recorded as the cursor moves from source to target location. This introduces transition delay. Also, unlike touch input, the user often overshoots or undershoots the target resulting in increased effective target width and error.}
  \label{fig:mousevisualization}
\end{figure}

\begin{figure}[!ht]
  \centering
  \includegraphics[width=3.0in]{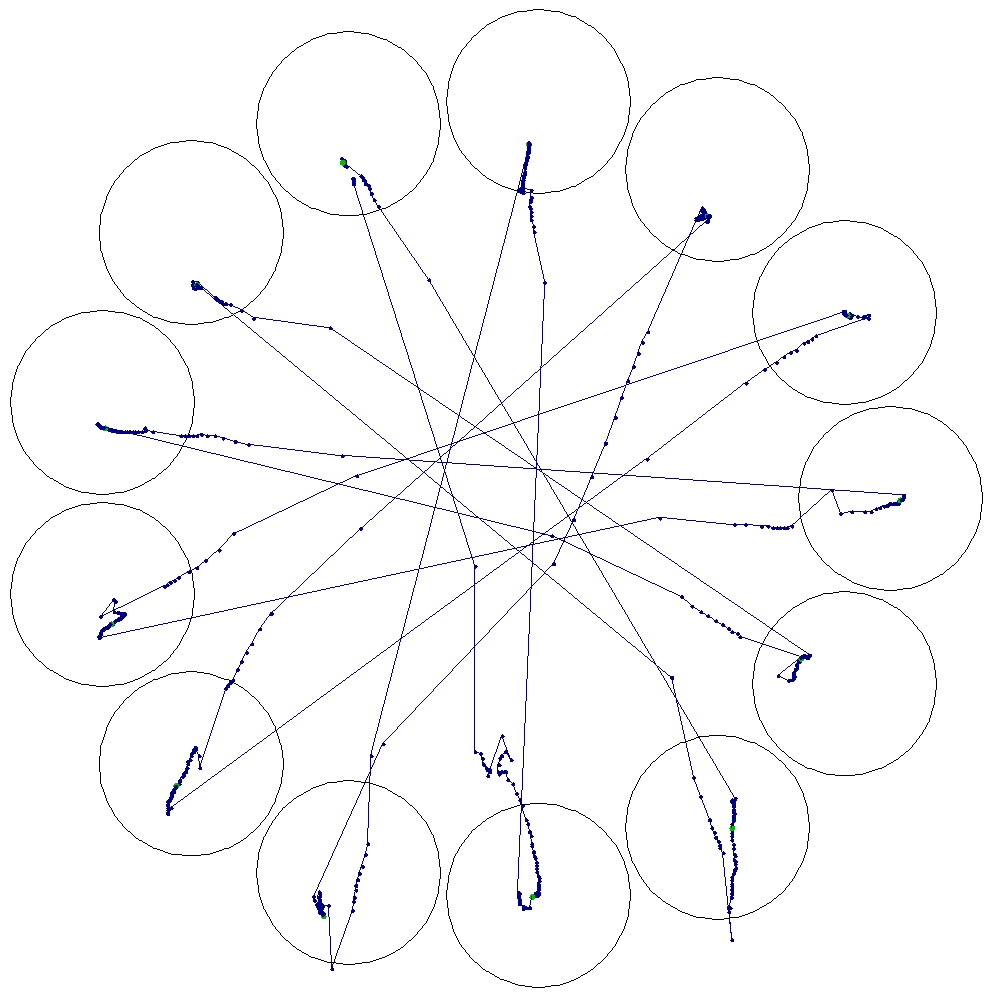}
  \caption{Gaze input cursor points visualization: compared to the mouse input there are only a few points recorded as the cursor moves from source to target location, this behavior is similar to the touch input. However, the maximum time is consumed in stabilizing the cursor (gaze) within the target, and selecting it. Hence, gaze has the highest movement time, effective target width, and lowest throughput than touch and mouse inputs.}
  \label{fig:gazevisualization}
\end{figure}

To reason, why does touch input perform the best, consider Figure~\ref{fig:touchvisualization}, Figure~\ref{fig:mousevisualization}, and Figure~\ref{fig:gazevisualization} that visualize the paths on the screen along which the mouse traversed during touch, mouse, and gaze inputs respectively.
The transition of input in subsequent point and select tasks as shown in Figure~\ref{fig:touchvisualization} for touch input clearly shows that the transition paths are well planned.
The user has greater control over how much to move her hand so that the target is not overshot and the selection is accurate.
Since the user hardly overshoots the target, there is minimal to no time consumed in correcting the transition path.
Furthermore, since the entire screen is within the view of the user, the user's hand can move quickly between the targets during selection.
These are the primary reasons that contribute to highest throughput and lowest movement time of touch input.


Next, we also observe that touch has the lowest error rate and effective target width.
As humans have better motor control over the movement of their hand, the user always hits inside the target, hence the lowest error.
Also, the touch-based selections are such that the user always selects the target at its center which leads to lower overshoot and undershoot values.
This behavior is unlike mouse and gaze inputs where users often overshoot or undershoot the target.
Hence, touch input has the lowest effective target width compared to mouse and gaze inputs.

Furthermore, considering why does gaze input have the lowest throughput, we can observe from Figure~\ref{fig:gazevisualization} that gaze input is indeed quicker in moving between the source to target location, similar to the touch input.
There are only a few cursor points along the path connecting the two targets.
However, when using gaze input, maximum time is consumed in stabilizing the cursor (gaze) within the target, and selecting it.
Hence, from Figure~\ref{fig:gazevisualization} we observe a cluster of points within the target. 
In addition, the higher effective target width results in lower index of difficulty.
Lower index of difficulty coupled with higher movement time results in lower throughput.
Hence, when using gaze input, adopting a border crossing strategy as the selection method will result in efficient interactions.

\section{Conclusion}
In this work, we conducted Fitts' Law evaluation of multimodal gaze and foot-based input by comparing it to touch input in a desktop environment.
Mouse input was used a baseline measure.
We expected that saccadic eye movements that allow pointing to be quickly moved from source to target location likely result in gaze input matching or even achieving higher throughput than direct input like touch.
However, results showed that while gaze was indeed quicker in moving between the source and target locations similar to touch input, maximum time was consumed in stabilizing the cursor inside the target and selecting it.
This resulted in gaze input achieving significantly lower throughput (2.55 bits/s) compared to touch (6.67 bits/s) and mouse inputs (3.8 bits/s).
Also, gaze input resulted in highest movement time (1.04 s) compared to touch (0.5 s) and mouse inputs (0.7 s).
Overall, with over 160\% higher throughput than gaze, touch input proved to be a superior input modality.

\bibliographystyle{acm-sigchi}
\balance{}
\bibliography{main.bbl}
\end{document}